\documentclass[twocolumn]{aastex631}

\usepackage{amsmath}
\graphicspath{{./}{figures/}}
\newcommand{\msol}{M_{\sun}}

\newcommand{\erg}{\rm erg}

\newcommand{\gcm}{\rm g~cm^{-3}}
\newcommand{\kbpb}{k_B/\rm baryon}

\newcommand{\qm}{\rm qm}
\newcommand{\on}{\rm on}
\newcommand{\fin}{\rm fin}

\begin{document}



\title{Phase-transition-induced collapse of proto-compact stars and its implication for supernova explosions}

\author[0000-0003-1842-8657]{Xu-Run Huang}
\affiliation{Department of Physics, the Chinese University of Hong Kong, Hong Kong S. A. R., China}

\author[0000-0001-6773-7830]{Shuai Zha}
\affiliation{Yunnan Observatories, Chinese Academy of Sciences (CAS), Kunming 650216, China}
\affiliation{International Centre of Supernovae, Yunnan Key Laboratory, Kunming 650216, China}

\correspondingauthor{Shuai Zha}
\email{zhashuai@ynao.ac.cn}

\author[0000-0002-1971-0403]{Ming-chung Chu}
\affiliation{Department of Physics, the Chinese University of Hong Kong, Hong Kong S. A. R., China}

\author[0000-0002-8228-796X]{Evan P. O'Connor}
\affiliation{The Oskar Klein Centre, Department of Astronomy, Stockholm University, AlbaNova, SE-106 91 Stockholm, Sweden}

\author[0000-0002-7444-0629]{Lie-Wen Chen}
\affiliation{School of Physics and Astronomy, Shanghai Key Laboratory for
Particle Physics and Cosmology, and Key Laboratory for Particle Astrophysics and Cosmology (MOE),
Shanghai Jiao Tong University, Shanghai 200240, China}

\begin{abstract}
A hadron-quark phase transition (PT) may trigger supernova explosions during stellar core collapse.
However, both success and failure have occurred in previous attempts to explode dying stars via this mechanism.
We systematically explore the outcomes of the PT-induced collapse of mock proto-compact stars (PCSs) 
with a constant entropy and lepton fraction, with spherically symmetric general relativistic hydrodynamic simulations and a controlled series of hybrid equations of state.
Our results reveal the qualitative dependence of successful and failed explosions on the PT and quark matter characteristics.
A small portion ($\sim\!0.04\%\!-\!1\%$) of the released binding energy $\Delta E_B$ transforms into the diagnostic explosion energy $E_{\rm exp,diag}$, which saturates at $\sim\!6\times10^{51}$\,erg near the black hole formation. 
Note that our $E_{\rm exp,diag}$ represents an upper limit of the final explosion energies in realistic supernova simulations.
We draw the phase diagrams \emph{indicative} of the possible fates of supernova explosions driven by hadron-quark PTs, where the control parameters are the onset density, energy gap of the PT, and the quark matter speed of sound. 
Our findings can 
guide further self-consistent investigations on PT-driven core-collapse supernovae and help identify hadron-quark PT-induced PCS collapse from future observations.
\end{abstract}


\section{Introduction} \label{sec:intro}

Massive stars die violently by the gravitational collapse of stellar cores and give birth to neutron stars (NSs) or stellar-mass black holes (BHs).
Sometimes, they are observed as Type II/Ibc supernovae and identified as core-collapse supernovae (CCSNe). 
The central engine of CCSN explosions has been under investigation for more than half a century but still not fully understood (see \citealt{1990RvMP...62..801B,2012ARNPS..62..407J,2020LRCA....6....3M,2021Natur.589...29B}  
for reviews).
Currently, two well-explored scenarios exist: the delayed neutrino heating mechanism \citep{1985ApJ...295...14B} which possibly produces most explosions with an energy up to a few $10^{51}$\,erg, and the magnetohydrodynamic mechanism \citep{1970ApJ...161..541L} which may be responsible for those extraordinary ``hypernovae" up to $\sim\!10^{52}$\,erg.
In addition, a strong hadron-quark phase transition (PT) occurring in the proto-compact star (PCS) may trigger successful CCSN explosions (see, e.g.,~\citealt{2009PhRvL.102h1101S}), even for very massive blue supergiant stars~\citep{2018NatAs...2..980F}.
 
Though free quarks may get deconfined from nucleons in the supernova interior \citep{1984PhRvD..30..272W}, the quantum chromodynamics (QCD) phase diagram in such extreme environments is quite unclear.
The existence of quark-gluon plasma has already been confirmed at relatively high temperatures and low densities via heavy-ion collisions in terrestrial laboratories~\citep{2005NuPhA.757..102A,2017Natur.548...62A}.
Recent progress in analyzing observations of massive pulsars favors the occurrence of quark deconfinement in their dense and cold cores~\citep{2023NatCo..14.8451A}.
Hadron-quark PTs can naturally occur in the PCS interior since the thermodynamic conditions realized throughout their evolution reside in the QCD phase diagram between that of heavy-ion collisions and cold NSs~\citep{2008RvMP...80.1455A,2018RPPh...81e6902B}.
Nevertheless, there is no current consensus on the exact properties of the PT \citep{2017RvMP...89a5007O,2021PhRvD.103b3001B}.

Previous efforts have explored the consequences of hadron-quark PTs in detailed CCSN simulations (see, e.g.,~\citealt{2013A&A...558A..50N,2020PhRvL.125e1102Z,2022MNRAS.516.2554J,2022ApJ...924...38K,2024ApJ...964..143K}).
By assuming a strong PT, successful explosions have been obtained even in spherical symmetry
~\citep{2009PhRvL.102h1101S,2018NatAs...2..980F}.
Simulations show that the massive PCS collapses to a small radius after the onset of PT and rebounds to form a second bounce shock.
Such a second collapse and bounce will normally be accompanied by the release of a millisecond neutrino burst \citep{2009PhRvL.102h1101S} and a distinct gravitational wave (GW) signal~\citep{2020PhRvL.125e1102Z}. 
These two signals provide unique signatures
of quark deconfinement in future galactic CCSNe~\citep{2013ApJ...778..164A}, and may also shed light on the elusive nature of neutrino~\citep{2022PhRvD.106j3007P}.
Along with the GW signals from isolated and merging NSs involving PTs~\citep{2006ApJ...639..382L,2019PhRvL.122f1102B,2019PhRvL.122f1101M,2023PhRvL.130i1404F}, future astrophysical observations can considerably enrich our understanding of the QCD phase diagram.

The outcomes of previous self-consistent simulations 
heavily depend on the properties of the employed hybrid equation of state (EoS) and the progenitor models.
A suite of 97 simulations in \citet{2022MNRAS.516.2554J} gives only 2 successful explosions, which suggests that the viability of PT-driven explosions remains ambiguous.
In this work, we carry out a systematic study on collapsing mock PCSs with a constant entropy and lepton fraction to explore the general relationship between the PT characteristics and outcomes. Our results present a qualitative framework for understanding the PT-driven explosion mechanism in realistic supernova simulations.
The paper is organized as follows. Section~\ref{sec:model} describes our simplified model, including the simulation setup and parametrized hybrid EoS. We also show the collapsing dynamics of the isentropic PCS by a representative case. Section~\ref{sec:systematism} demonstrates the systematic relationship between the outcomes and PT characteristics, and its implication for supernova explosions. We conclude in Section~\ref{sec:sum}.

\section{Modeling the PT-driven explosion} \label{sec:model}

Canonical CCSN simulations feature the production of a PCS mainly composed of hadrons, after the implosion of the stellar core.
A shock wave generated during the first core bounce propagates out and stalls at $\sim$100-200\,km.
The system enters a period of mass accretion onto the surface of the nascent PCS, which grows in mass and contracts. 
The PCS central density eventually exceeds several times the nuclear saturation density, $\rho_0 \simeq 2.6\times 10^{14}$\,g~cm$^{-3}$ or $n_{0} = 0.155\,\rm{fm^{-3}}$.
Note that the PCS matter is too dense for photons to decouple, and even neutrinos get trapped in the inner core (with densities higher than $\sim\!10^{12}$\,g~cm$^{-3}$).  
The PCS may undergo a hadron-quark PT at this stage and collapse into a hot hybrid star with a smaller radius.
Previous simulations have shown that a second core bounce, induced by the stiffening of quark matter EoS, can generate a new shock wave and lead to a successful explosion \citep{2009PhRvL.102h1101S,2018NatAs...2..980F,2020PhRvL.125e1102Z}.

In this work, we concentrate on the 
collapse and second bounce phase of the nascent PCS. 
As neutrinos get mostly trapped in the interior of PCS,
we assume a $\beta$-equilibrium with constant electron lepton number fraction of $Y_L = 0.3$ for the nuclear EoS.
In previous surveys of the progenitor dependence \citep{2020ApJ...894....4D,2021ApJ...911...74Z}, the specific entropy turns out to be almost uniform in the outer layer of a PCS at the juncture of collapsing into BH, e.g., $s \simeq 4.7\,\kbpb$ and $2.7\,\kbpb$ in the gravitational mass coordinates of $[1.0\,M_\sun,2.0\,M_\sun]$ and $[0.5\,M_\sun,2.0\,M_\sun]$ for a high and low compactness pre-supernova progenitor, respectively (see Figures 5 and 6 in \citet{2020ApJ...894....4D}).  
They further find that the most common entropies\footnote{The most common entropy of a PCS is defined as the specific entropy weighted by mass (see Equation (4) in \citet{2021ApJ...911...74Z}), which avoids the disguisement effect from both the central cold region of PCS and the hot region right below the accretion shock.} of PCSs are typically in the range of $3\!\sim\!5\,\kbpb$ during a few seconds postbounce. 
To capture the collapse moment in actual simulations, we build up isentropic PCSs (described in Section~\ref{subsec:EoS}) 
with $s\sim3-4\,\kbpb$ to mimic the conditions right before their collapse.
With these simplifications, we perform spherically symmetric simulations of PCSs with the general-relativistic hydrodynamic code {GR1D}~\citep{2010CQGra..27k4103O,2015ApJS..219...24O}.
The grid size is uniform of 100\,m in the innermost $50\,\text{km}$, and becomes logarithmically increasing up to $10^4$\,km with 500 more zones. Outside the PCSs ($\sim$20-30\,km), we set an atmosphere with a constant density of $1\,\gcm$ which is sufficiently low to avoid affecting the shock dynamics.

\subsection{Hadron-Quark Hybrid EoS}\label{subsec:EoS}

A series of hadron-quark hybrid EoSs are constructed with a first-order PT, from the {SFHo} relativistic mean field model~\citep{2013ApJ...774...17S}.
The hadronic part is described by the isentropic EoS with a specific entropy $s = 4\,\kbpb$.
Quark matter is parametrized in terms of three quantities: the pressure of the transition from nuclear matter to quark matter ($P_{\on}$), the energy gap of the PT ($\Delta \varepsilon$) and the sound speed in the quark matter ($c_{\qm}$). This constant-sound-speed (CSS) parameterization~\citep{2013PhRvD..88h3013A} can be written as:
\begin{equation}
    \varepsilon(P) = 
    \begin{cases}
    \varepsilon_{\rm{NM}} (P) & P<P_{\on} \\
    \varepsilon_{\on} (P_{\on}) + \Delta\varepsilon + (P-P_{\on})\beta & P>P_{\on}
    \end{cases},
\label{eq:cssEoS}
\end{equation}
where $\beta = c^2/c^2_{\qm}$ and $\varepsilon_{\on} (P_{\on})$ is the total energy density of nuclear EoS $\varepsilon_{\rm{NM}} (P)$ at the transition density $\rho_{\on}$.
The Maxwell construction~\citep{1992PhRvD..46.1274G,2001PhR...342..393G} 
is employed for the PT, i.e., the pressures and baryon chemical potentials\footnote{We take neutron chemical potentials for the hadronic matter.} are equilibrated across the PT region.
The effect of temperature variation is neglected across the PT region.
The baryonic density $\rho$ for quark matter is derived from $P$ and $\varepsilon$ following the Appendix of \citet{2013PhRvD..88h3013A}. 
Explicitly, the hybrid EoS is given by
\begin{equation}
P(\rho) = 
    \begin{cases}
    P_{\rm{NM}} (\rho) & \rho\le\rho_{\on} \\
    P_{\on} & \rho_{\on}<\rho\le\rho_{\fin} \\
    a\left(\dfrac{\rho/m_u }{a (1+\beta)}\right)^{1+\frac{1}{\beta}}-b & \rho_{\fin}<\rho
    \end{cases},
    \label{eq:eosPrho}
\end{equation}
where $m_u$ is the atomic mass unit. Using the relation $\varepsilon = n\mu - P$ where $n$ and $\mu$ are baryon number density and baryon chemical potential respectively, the transition density to pure quark matter (the end of the mixed phase) is given by 
\begin{equation}
    \rho_{\fin} = m_u(\varepsilon_{\on} + \Delta\varepsilon + P_{\on})/\mu_{\on},
\end{equation}
with $\mu_{\on}$ being the neutron chemical potential at $\rho_{\on}$. $a$ and $b$ are determined by the following equations
\begin{equation}
\begin{cases}
    P_{\on} = a \mu_{\on}^{1+\beta} - b \\
    \varepsilon_{\on}+\Delta\varepsilon = \beta P_{\on} + (1+\beta)b
\end{cases}.
\end{equation}

Note that the CSS model was originally devised to describe quark matter at zero temperature, and contains no composition and temperature dependence.
Previous surveys show that this CSS parametrization can accurately characterize some cold EoSs of quark matter based on the thermodynamic bag model~\citep{2016PhRvD..94j3008H} and the Nambu–Jona-Lasinio (NJL) model~\citep{2016PhRvC..93d5812R}.
Nevertheless, quark matter would have a much weaker entropy-dependence than hadronic matter in the density range of interest, e.g., in the plane of $P$ vs. $\rho$ for the hybrid EoSs in \citet{2021PhRvD.103b3001B}, only an upward shift of $\lesssim\!10\%$ is found when varying $s$ from $3\,\kbpb$ to $5\,\kbpb$ in the quark phase, while it can reach $\sim\!190\%$ in the hadronic phase.
We thus use the CSS model to capture the main features of the isentropic quark matter.
The Maxwell construction assumes that the surface tension is high enough to ensure that quark and hadronic potions are well separated in the mixed phase, which is favored by some realistic quark matter models~\citep{2013PhRvC..88d5803L,2020PhRvD.102h3003R}.

We restore approximately the thermal effect induced by shock heating 
via an ``ideal-gas" ansatz (see, e.g., \citealt{1993A&A...268..360J,2010PhRvD..82h4043B,2019PhRvL.122f1102B}), 
with a thermal index $\Gamma_{\rm th}$ mimicking the behavior of full EoSs ~\citep{2010PhRvD..82h4043B,2013ApJ...774...17S,2023PhRvD.108f3032B}. 
Explicitly, $\Gamma_{\rm th}$ is given by
\begin{equation}
    \Gamma_{\rm th} = 
    \begin{cases}
        4/3 & \log(\rho) < 11 \\
        4/3 + 5(\log(\rho)-11)/36 & 11 \le \log(\rho) < 14 \\
        7/4 &  14 \le \log (\rho) < \log (\rho_{\on}) \\
        1 &  \rho_{\on} \le \rho < \rho_{\rm fin} \\
        4/3 & \rho_{\rm fin} \le \rho
    \end{cases}.
    \label{eq:Gammath}
\end{equation}
This thermal component becomes relevant only for 
extra internal energies beyond the barotropic EoS (Equations~\eqref{eq:cssEoS} and \eqref{eq:eosPrho}).

The caveat is that the Maxwell construction in a realistic model is defined at a constant temperature, which normally leads to a smooth increase of the pressure across the PT at a constant entropy (e.g., the hybrid EoSs in \citet{2021PhRvD.103b3001B}). The flat slope of the pressure in our isentropic model would cause the sharpest drop immediately on the polytropic index \citep{2023PhRvD.107g4013C}, which facilitates the collapse of PCSs.
It also implies the vanishing of sound speed during the PT. 
This precludes the Jacobian matrix of the system of general relativistic hydrodynamic equations to obtain real and distinct eigenvalues (see Equations (20-24) in \citet{1996ApJ...462..839R}), and terminates further numerical evolution of hydrodynamic equations.
To avoid this problem, we artificially raise the value of sound speed $c_s$ in simulations by a tiny number over the PT region, e.g., $h c_s^2 = 10^{-50} c^2$ with $h$ being the specific enthalpy.
In addition, linear interpolation is implemented for the energy density in the PT region.

\begin{figure}[t!]
 \centering
 \includegraphics[width=1\columnwidth]{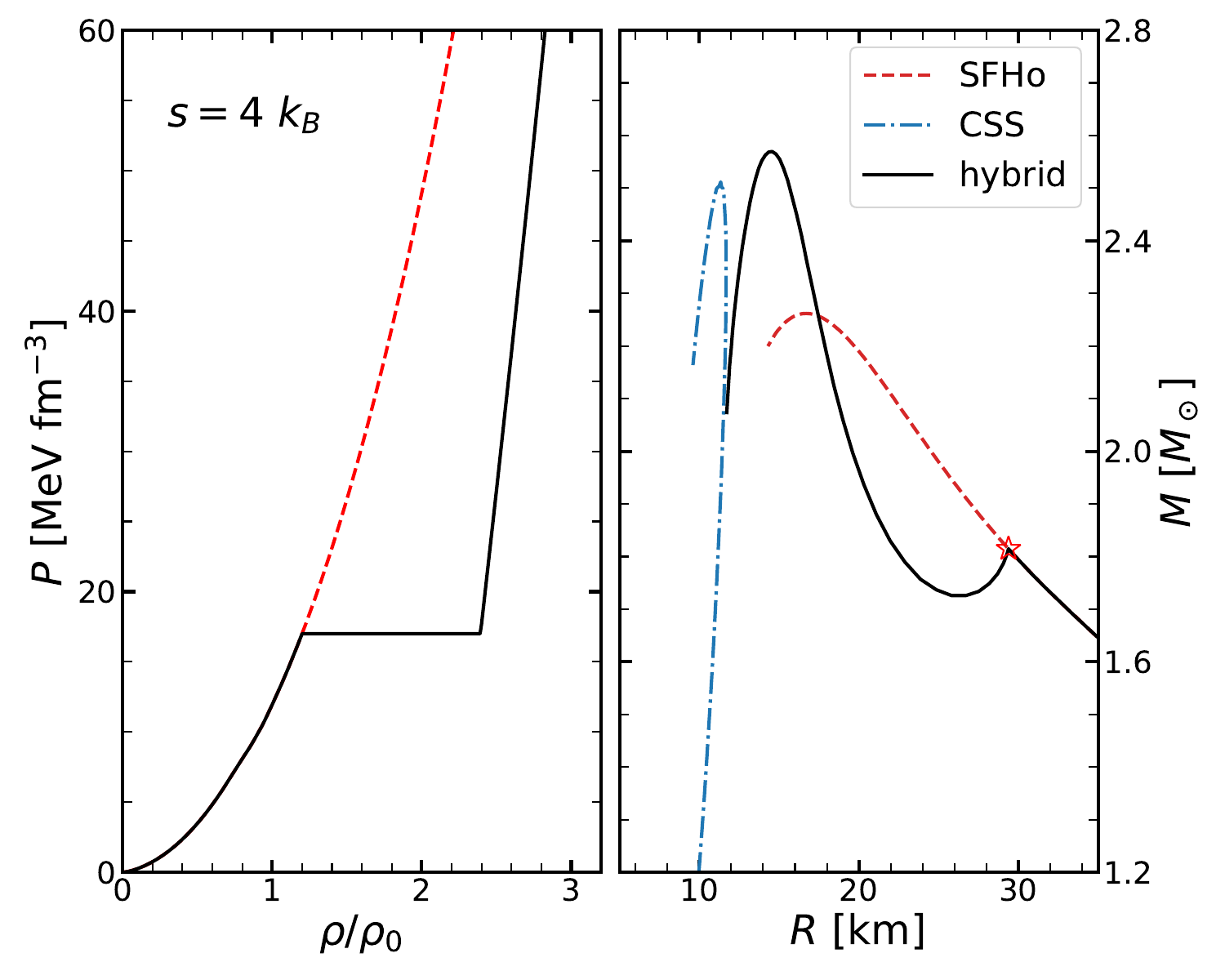}
 \caption{A representative hybrid EoS (solid lines) constructed from the {SFHo} EoS (dashed lines) at $s = 4\,\kbpb$ with an onset density of phase transition at $\rho_{\on} = 1.2\,\rho_0$. 
 {\tt \string Left panel:} Pressure $P$ versus rest-mass density $\rho$. {\tt \string Right panel:} The M-R relations of hot compact stars with M being the gravitational mass. The dashed, dot-dashed, and solid curves correspond to hadronic, quark, and hybrid stars, respectively. The red star marks the onset mass of hadron-quark PT, i.e., $M_{\on} \simeq 1.8\,\msol$. }
\label{fig:EoS_MR}
\end{figure}
Figure~\ref{fig:EoS_MR} shows a typical model with $\rho_{\on} = 1.2\,\rho_0$, $\Delta\varepsilon/\varepsilon_{\on} = 0.6$ and $c^2_{\qm} = 2/3\,c^2$ which has a PT region from 1.2$\,\rho_0$ to 2.4$\,\rho_0$.
The squared sound speed jumps from $\sim\!0.15 \,c^2$ for hadronic matter to 0 in the PT region and leaps to $2/3 \,c^2$ for quark matter.
We also plot the mass-radius (M-R) relations of hot compact stars 
in the right panel.
The M-R curve for the hybrid EoS has two extreme masses, i.e., the most massive hybrid star ($M_{3,\max} \simeq 2.6\,\msol$) and purely hadronic  
star ($M_{2,\max} = M_{\on} \simeq 1.8\msol$) connected by a stable and unstable branch, while those for pure hadron and quark matter have only one extreme mass each.
This hybrid model forms the so-called ``disconnected" third-family topology  \citep{2013PhRvD..88h3013A,2016PhRvD..94j3008H}.
Previous analyses on hot and cold hybrid stars indicate that the existence of a third family is related to PT-driven CCSN explosions, especially for the hybrid EoSs with a high $\Delta\varepsilon/\epsilon_{\on}$ and low $\rho_{\on}$~\citep{2016PhRvD..94j3001H,2016PhRvD..94j3008H}. Our simulations further confirm such an expectation (see Figures~\ref{fig:PhaseDiag1} and \ref{fig:PhaseDiag2}).

\subsection{Proto-Compact Star Evolution-A representative Case}\label{subsec:reprcase}

\begin{figure*}
  \centering
  \includegraphics[width=.32\linewidth]{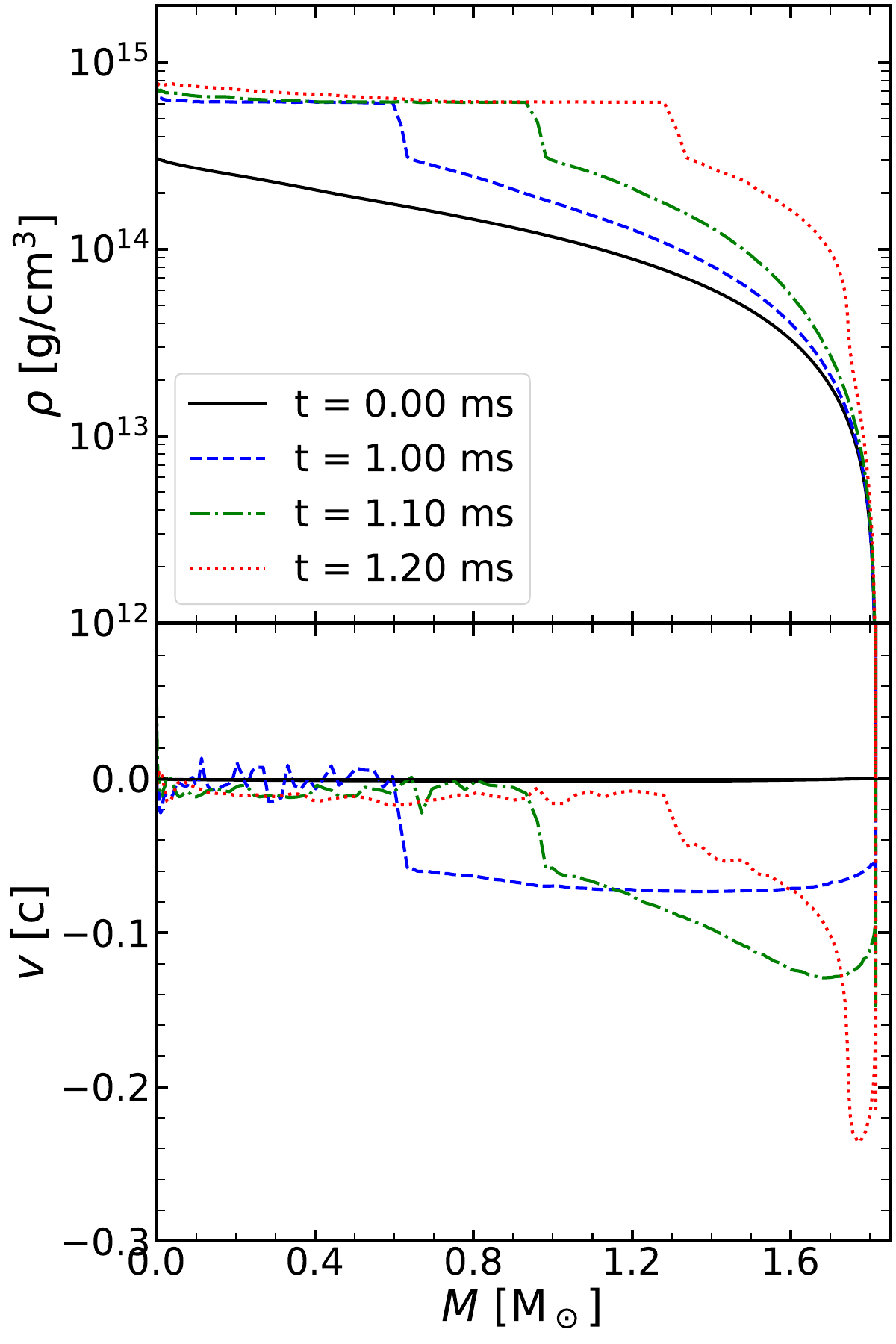}
  \includegraphics[width=.32\linewidth]{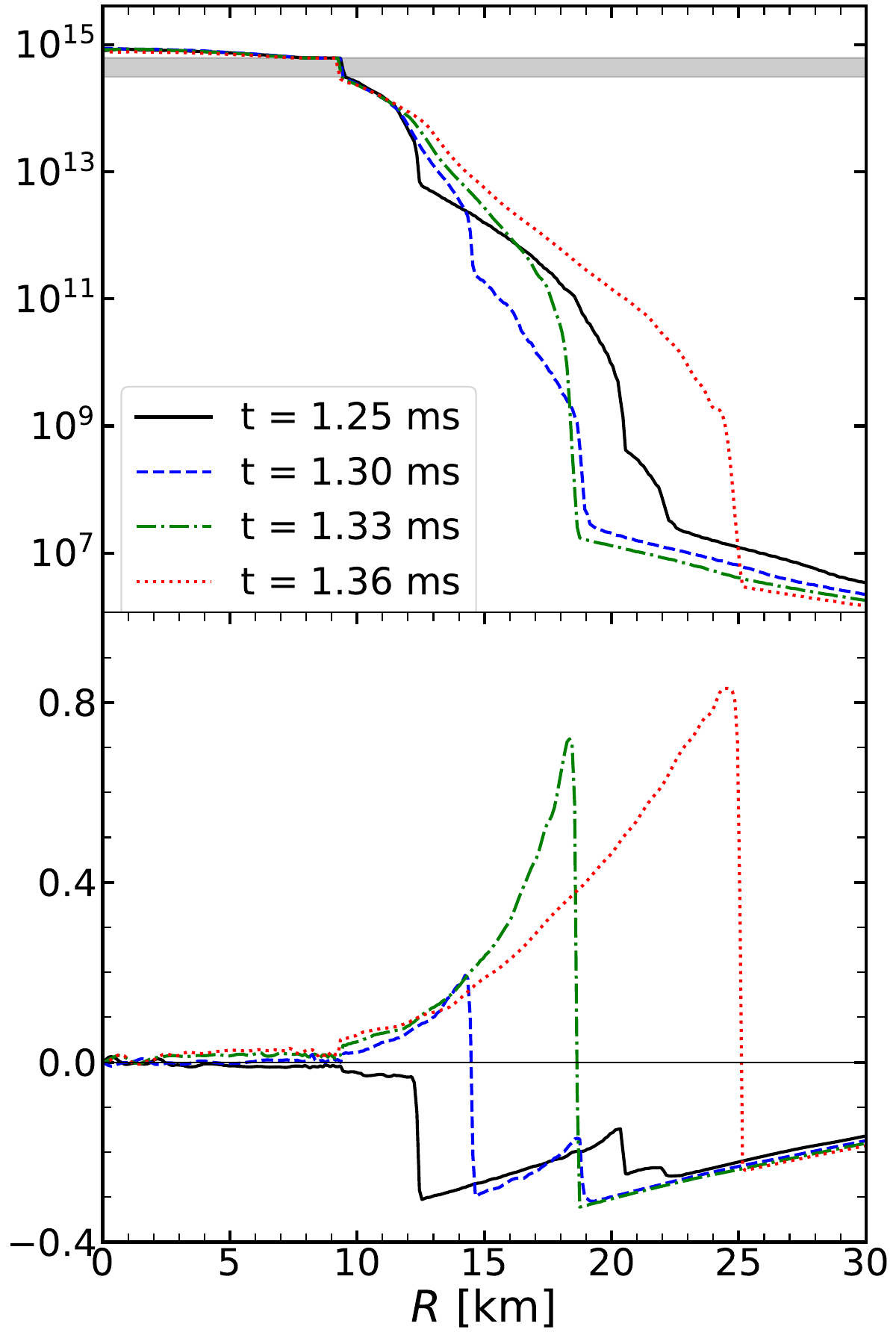}
  \includegraphics[width=.32\linewidth]{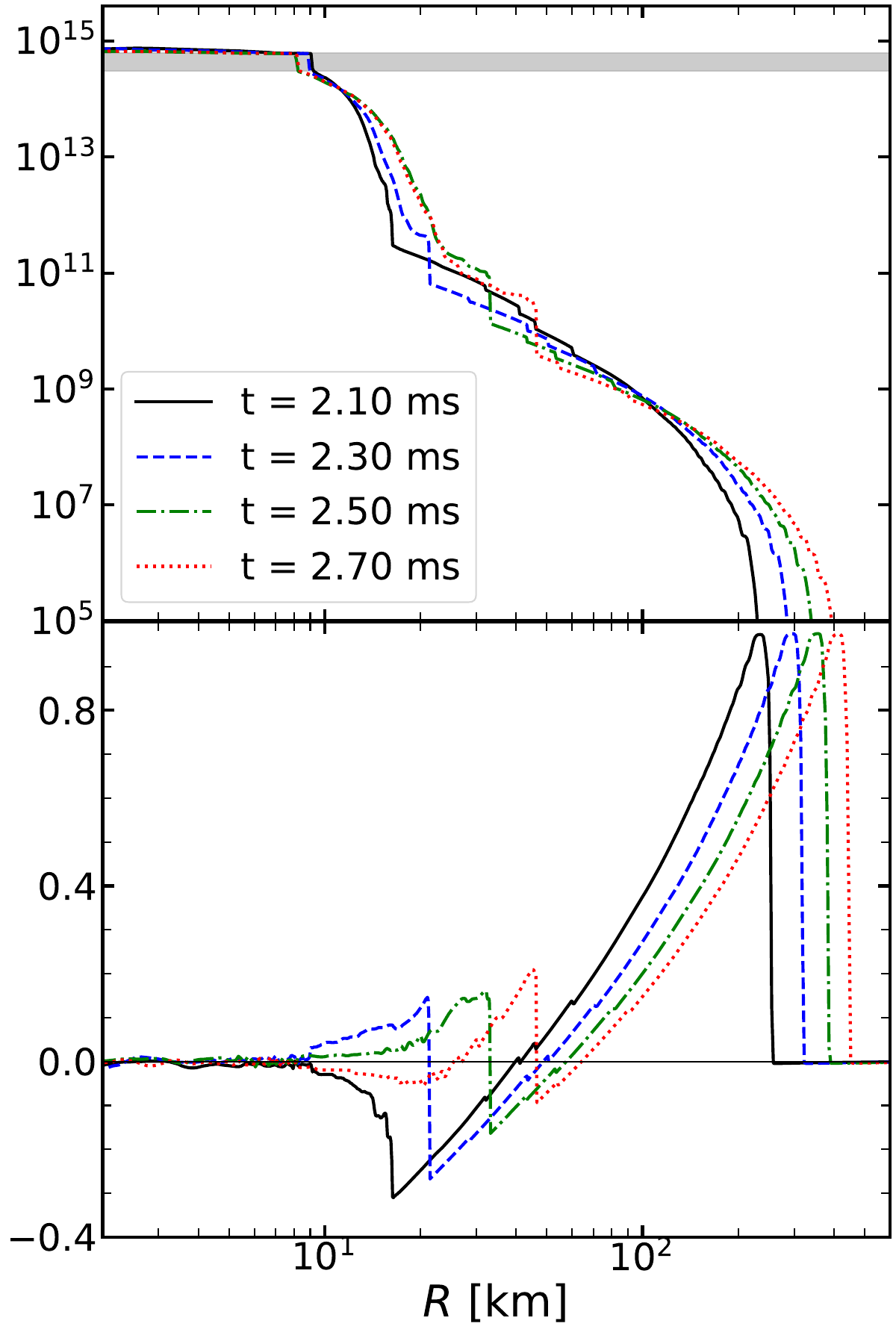}
\caption{Radial profiles for the representative model at selected evolution times, 
showing the density $\rho$ (upper panels) and radial speed $v$ in units of the speed of light $c$ (lower panels). The selected quantities are displayed as functions of the enclosed gravitational mass (left), as well as the radius (middle, right). The gray bands in the upper panels mark the range of the PT.}
\label{fig:xgEvo}
\end{figure*}
The hydrodynamic simulation of a PCS begins at its onset mass of PT, e.g., $\sim\!1.8\,\msol$ in the above model.
Its evolution is illustrated in Figure~\ref{fig:xgEvo}.
Initially, we perturb the PCS by adding a radial infall velocity as
\begin{equation}
    v(x) = 
    \begin{cases}
        - 2x v_{\rm max} & x < \frac{1}{2} \\
        -v_{\rm max} + 4(x - \dfrac{1}{2})v_{\rm max} &  \frac{1}{2} \le x < \frac{3}{4} \\
        0 &    x \ge \frac{3}{4}
    \end{cases},
    \label{eq:pertbVelocity}
\end{equation}
where $x = r/R_{\rm pcs}$ with $R_{\rm pcs}$ being the PCS radius and $v_{\rm max}/c = 0.2\%$.
The PCS becomes gravitationally unstable and starts to collapse.
Within the first ms, it contracts slowly with the formation and growth of a quark core.
The quark core mass reaches $\sim\!0.6\,\msol$ while the maximum infall velocity accelerates to a few percent of $c$.
We remark that both the conditions of PCS at this moment and the following evolution look quite similar to the PCS collapse in the realistic CCSN simulations (see, e.g., Figure 1 in \citet{2024ApJ...964..143K}).
In the following 0.2 ms,  
the hadronic shell collapses violently and the infall velocity exceeds $20\%$ of $c$ (see the left panel in Figure~\ref{fig:xgEvo}).
The quark core remains quasi-static and expands swiftly into $\sim\!1.3\,\msol$.
The collapsing inner hadronic shell overshoots the highly incompressible quark core 
and bounces back.
The outgoing pressure waves 
steepen into a shock front at the transition to the supersonically infalling shell.
Such an evolution turns into the sharp velocity gradient at the shock front (see the middle panel in Figure~\ref{fig:xgEvo}).
The shock then expands rapidly through the steep PCS surface and results in matter outflows with velocities exceeding $0.8\,c$.
After that, a secondary shock wave is generated and propagates outward with reduced speeds (see the right panel in Figure~\ref{fig:xgEvo}).

After the PCS reaches its final hydrostatic equilibrium (e.g., $>\!7$\,ms), we make a conjecture about the final explosion energy $E_{\rm exp,diag}$ by integrating the reduced binding energy $e_{\rm bind}$ over the region where $e_{\rm bind}$ is positive (see Equations (3-4) in \citealt{2012ApJ...756...84M}). The above model produces $E_{\rm exp,diag} \simeq 1.0$\,B ($1\,\rm{B} = 10^{51}\,\erg$).

\section{A systematic study in the QCD Phase Transition Driven Explosion} \label{sec:systematism}

In this section, we systematically survey how the fate of PT-induced post-bounce dynamics depends on the PT characteristics.

\subsection{Phase Transition Dependence and its Interpretation}\label{subsec:interp}

\begin{figure*}[htb!]
 \centering
 \includegraphics[width=0.95\linewidth]{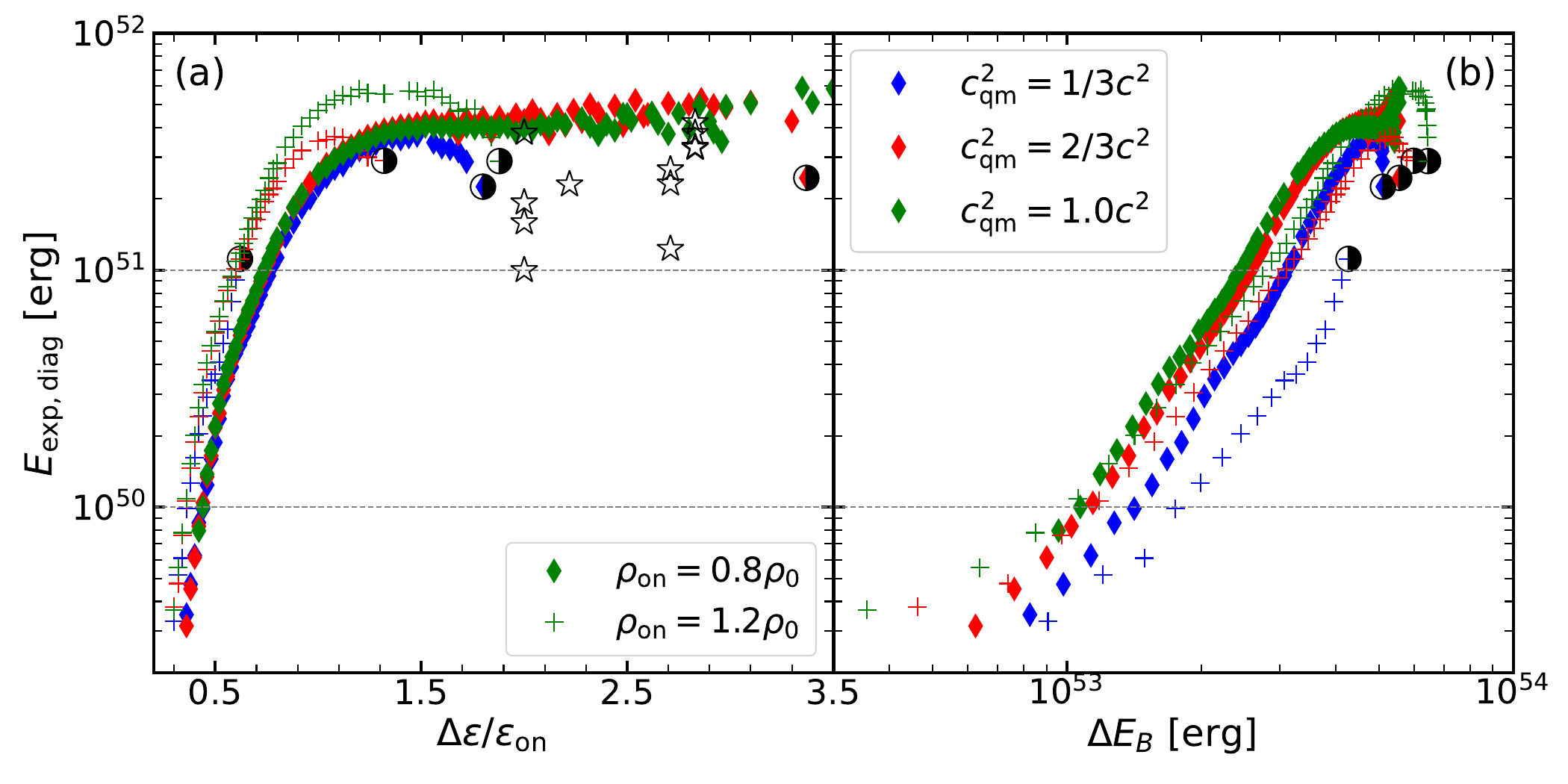}
 \caption{Dependence of the final $E_{\rm exp,diag}$ on parameters of hadron-quark PT (a) and $\Delta E_B$ (b) (Equation~\eqref{eq:Ebinding}). Results are displayed for different $c^2_{\qm}$ (colors) and $\rho_{\on}$ (markers). The half-filled black circles mark the data point before prompt BH formation, except that of the $\rho_{\on} = 0.8\,\rho_0$ series (green diamonds), which falls outside this range.
 The black unfilled stars denote the explosion models reported in \citet{2024ApJ...964..143K}. Note that $E_{\rm exp,diag}$ in our idealized models represents the upper limit of explosion energies in realistic simulations.} 
\label{fig:Analy}
\end{figure*}
The PT dependence of the final $E_{\rm exp,diag}$ is illustrated in Figure~\ref{fig:Analy} (a).
Firstly, it is evident that $E_{\rm exp,diag}$ increases as $\Delta\varepsilon$ increases, but saturates until a certain large $\Delta\varepsilon$ beyond which explosion fails and prompt BH formation occurs.
In the BH formation cases, the central density rises abruptly to exceed about $15\!\sim\!20\,\rho_0$ and the central lapse function drops below a few $10^{-2}$. 
$E_{\rm exp,diag}$ is suppressed near the critical values of $\Delta\varepsilon$ right before prompt BH formation, marked by the half-filled black circles.
$E_{\rm exp,diag}$ for $\rho_{\on} = 1.2\,\rho_0$ (plus symbols) shifts systematically upward when compared to that of $\rho_{\on} = 0.8\,\rho_0$ (diamond symbols). $E_{\rm exp,diag}$ turns out to be approximately insensitive to $c^2_{\qm}$ except for the region close to prompt BH formation. 
This can be understood as a softer quark matter reduces the power of bounce but leads to a more compact hybrid star (smaller radius) which increases the released binding energy.

Secondly, both $\rho_{\on}$ and $c^2_{\qm}$ have a great impact on the critical $\Delta\varepsilon/\varepsilon_{\on}$ for prompt BH formation. A larger $\rho_{\on}$ decreases it, since a larger $M_{2,\max}$ is reached which facilitates the BH formation. However, a larger $c^2_{\qm}$ hinders it, because a stiffer quark matter increases $M_{3,\max}$ which impedes the further collapse of the PCS. 
We also find that the condition of BH formation is reached in the center before the gravitational mass of the PCS exceeds $M_{3,\max}$ by a deficit of $\sim5\%$-$19\%$.
Eventually, $E_{\rm exp,diag}$ reaches an upper limit at a moderate $\rho_{\on}$ and a large $c^2_{\qm}$, e.g., $E_{\rm exp,diag} \lesssim 6\,\rm{B}$ at $\rho_{\on} \simeq 1.1\,\rho_0$ and $c^2_{\qm} = 1.0\,c^2$. 

Thirdly, the explosion models obtained by full simulations\footnote{Quark matter appears off-center earlier than the collapse moment in detailed simulations due to the entropy-dependent PT, while it first appears in the center in our simulations. After the collapse moment, which is defined as when the PCS infall velocities reach a few percent of $c$, the evolution is similar in both cases.}, reported in \citet{2024ApJ...964..143K}, are also included in Figure~\ref{fig:Analy} (a) (black unfilled stars).
The onset densities of their hybrid EoSs are $\sim\!\rho_0$ when $s \simeq 4\,\kbpb$ and $Y_L = 0.3$. 
Their EoSs include extra repulsive interactions~\citep{2021PhRvD.103b3001B}, 
which results in a larger quark core and smaller differences in the gravitational binding energy of a hadronic core and quark core.
This reduces the released binding energy~\footnote{The released binding energy, which is estimated according to the M-R relations of isentropic compact stars at $s=4\,\kbpb$, is reduced by a factor of $\sim\!5$, when compared to the EoSs with similar parameters in our models.} and $E_{\rm exp,diag}$.
Note that their $E_{\rm exp,diag}$ also 
contain the contributions from the stellar envelopes of progenitors.
In practice, $E_{\rm exp,diag}$ would be shifted down by about $0.1\!\sim\!0.9$\,B if taking the envelope into account~\citep{2016ApJ...818..123B}.
Such a deficit is currently not considered in our models. 
Nevertheless, several different energies are reported for a single hybrid EoS in their work, due to different progenitor stars.
This phenomenon can be understood by the dependence on the compactness of the progenitor \citep{2021ApJ...911...74Z}. 
Generally speaking, the PCS from a less compact progenitor evolves with a smaller value of the entropy and tends to release less explosion energy in the PT-induced collapse.
Another issue is that our flat slopes of the pressures across the PT regions should give upper limits on the explosion energies, compared to the increasing slopes in realistic EoSs.
After considering the above effects, the detailed explosion models are fairly compatible with the predictions of our simplified models.

To draw a quantitative view of the above PT dependence, we estimate the binding energy released in the collapse of the PCS by 
\begin{equation}
    \Delta E_B = M_{2,\max} - M_{c,t}.
    \label{eq:Ebinding}
\end{equation}
The definition of $M_{c,t}$ is based on the existence of a third family topology in the M-R curve (cf. Figure~\ref{fig:EoS_MR}),
i.e., a hybrid twin star of $M_{2,\max}$ resides on the inner stable branch. A core of this hybrid twin star is extracted with the same baryonic mass as the hadronic twin star $M_{2,\max}$.
$M_{c,t}$ is defined as the gravitational mass of the extracted core.
Note that a larger $\Delta \varepsilon$ will normally result in a larger $\Delta E_B$.

We plot $E_{\rm exp,diag}$ as a function of $\Delta E_B$ in Figure~\ref{fig:Analy} (b). 
Firstly, considering one specific series of models (e.g., red diamonds), 
$\log (E_{\rm exp,diag})$ is linearly proportional to $\log (\Delta E_B)$, e.g., $\log (E_{\rm exp,diag}) \simeq 2.688 \log (\Delta E_B) - 92.587$. 
It also suggests that only a small part of $\Delta E_B$ transforms into $E_{\rm exp,diag}$, and the transformation efficiency has a similar linear dependence on $\log (\Delta E_B)$. 
The transformation efficiency rises roughly from 0.05\% to 0.9\% with the growth of $\Delta E_B$. 
$E_{\rm exp,diag}$ saturates at large $\Delta E_B$. $\Delta E_B$ also tends to saturate at large $\Delta\varepsilon$. This dual effect explains the aforementioned saturation of $E_{\rm exp,diag}$ at large $\Delta\varepsilon$ in Figure~\ref{fig:Analy} (a).
In our simulations, $E_{\rm exp,diag}$ is ultimately produced by the first PT-induced bounce shock for models with large $\Delta\varepsilon$.
Consequently, this bounce will get firstly strengthened and then saturated as $\Delta\varepsilon$ increases. 
A suppression effect also emerges before the boundary of prompt BH formation (e.g., blue diamonds).
However, there is a caveat in estimating $E_{\rm exp,diag}$ for cases close to BH formation as GR1D cannot continue past this point due to the choice of slicing \citep{2010CQGra..27k4103O}.

Secondly, at $\rho_{\on} = 1.2\,\rho_0$, models on the less energetic side (e.g., red pluses in $\Delta E_B < 10^{53}\,\erg$) noticeably deviate from the 
linear-dependence of $\log (E_{\rm exp,diag})$ vs. $\log (\Delta E_B)$.
We find that these models are typically located near the vanishing boundary of twin stars in the phase diagram for the M-R relation of compact stars (cf. Figure 3 in \citealt{2013PhRvD..88h3013A}).
This collective deviation is attributed to the boundary effect which suppresses $\Delta E_B$.

Thirdly, $E_{\rm exp,diag}$ for $c^2_{\qm}=2/3\,c^2$ (red diamonds in Figure~\ref{fig:Analy} (b)) shifts systematically upward when compared to that of $c^2_{\qm}=1/3\,c^2$ (blue diamonds in Figure~\ref{fig:Analy} (b)). 
At a fixed $\rho_{\on}$, such upward trends are universal among different $c^2_{\qm}$.
It also means that the transformation efficiency from $\Delta E_B$ to $E_{\rm exp,diag}$ is increased for a large $c^2_{\qm}$.
Actually, at a fixed $\Delta \varepsilon$, a large $c^2_{\qm}$ reduces $\Delta E_B$ noticeably, especially from $1/3\,c^2$ to $2/3\,c^2$, but leaves relatively small influence on $E_{\rm exp,diag}$.
As a result, the above improvement from $c^2_{\qm}$ occurs and tends to saturate after $2/3\,c^2$.
Finally, a large $\rho_{\on}$ can slightly reduce the slope between $\log (E_{\rm exp,diag})$ and $\log (\Delta E_B)$ at a fixed $c^2_{\qm}$.

\subsection{Fate of CCSNe with QCD Phase Transition}\label{subsec:fate}

Based on the results of our idealized models, we provide an indicative framework for the possible outcomes of realistic CCSN simulations.
A collapse of the PCS with $E_{\rm exp,diag} > 0.1\,\rm B$ is assumed to  
result in a successful SN explosion. 
This assumption is based on the fact that the explosion energy normally ranges from 0.1\,B to 5\,B for Type II SN light curves~\citep{2015ApJ...806..225P,2017ApJ...841..127M,2019ApJ...879....3G,2022ApJ...934...67B}.
We also regard 1\,B as another benchmark energy, since this is the magnitude of the estimated explosion energy of SN 1987A \citep{2014ApJ...783..125H}.

\begin{figure}[t!]
 \centering
 \includegraphics[width=1\columnwidth]{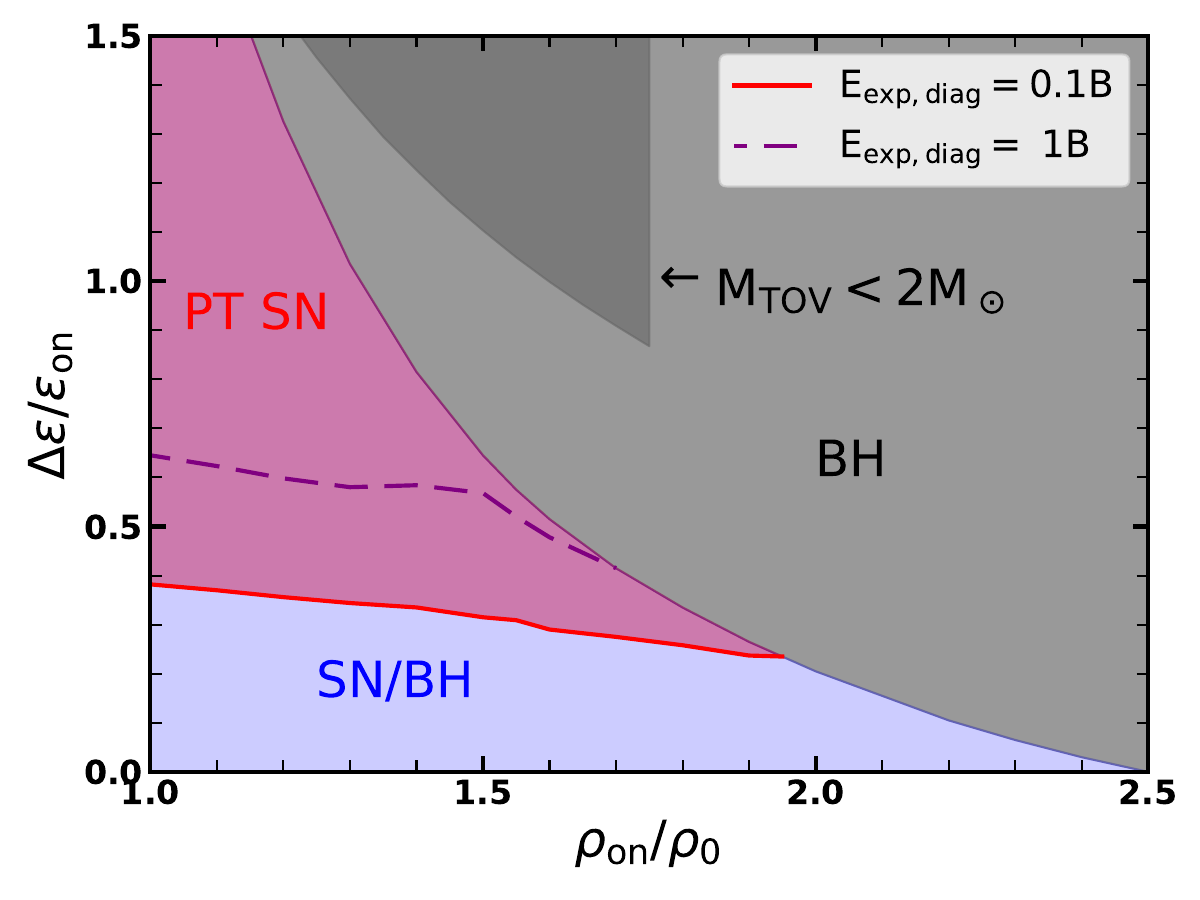}
 \caption{Phase diagram for the possible fates of PT-induced collapse of mock PCSs. The control parameters are $\rho_{\on}$ and $\Delta \varepsilon$ with $c^2_{\qm}=2/3\,c^2$ and $s = 4\,\kbpb$. 
 The parameter space for $E_{\rm exp,diag} <(>) 0.1\,\rm B$ is labeled as ``SN/BH" (``PT SN") and colored blue (red). 
 The light gray shaded region indicates prompt BH formation. 
 The dark region is excluded by the observed 2\,$M_\sun$ pulsars. }
\label{fig:PhaseDiag1}
\end{figure}
Figure~\ref{fig:PhaseDiag1} demonstrates the possible outcomes of CCSN simulations, according to $E_{\rm exp,diag}$ in the collapse of PCS.
$c^2_{\qm}$ is currently fixed at $2/3\,c^2$.
For a weak PT (blue), the second bounce is normally weak as well, e.g., $E_{\rm exp,diag} < 0.1\,\rm B$.
The post-bounce evolution should still be dominated by the delayed neutrino heating in the gain region, and it subsequently ends up as a canonical neutrino-driven explosion or delayed BH formation.
It is generally the latter scenario in spherical symmetry~\citep{2021ApJ...911...74Z,2022MNRAS.516.2554J,2024ApJ...964..143K}.
In contrast, an energetic second bounce shock can be generated for a strong PT (red). It takes over and merges with the stalled accretion shock.
This can potentially unbind the stellar envelope if the mass accretion rate is not too high, even in spherical symmetry \citep{2009PhRvL.102h1101S,2018NatAs...2..980F}.
The critical $\Delta \varepsilon/\varepsilon_{\on}$ between neutrino-driven and PT-induced explosions is approximately proportional to $A-B \rho_{\on}$, with $A$ and $B$ both positive. 
This can be well explained by the effect of $\rho_{\on}$ revealed in Figure~\ref{fig:Analy}.
We also display the critical line of $E_{\rm exp,diag} = 1\,\rm B$.
As expected, it is generally parallel to the above boundary, except for a bump near the region of prompt BH formation.

A strong PT launching at a high $\rho_{\on}$ (gray region) will lead to a prompt collapse into BH in our simulations.
Particularly interesting is that the collapse of PCS will inevitably result in a prompt BH formation if the PT launches at a density above $\sim\!2.5\,\rho_0$ (i.e., $M_{\on}\!\sim\!2.15\,\msol$).
This potentially renders a way to constrain the PT, by combining the initial mass function for precollapse stars and observed compact remnants of CCSNe. 
We also show the region (dark) where the maximum mass predicted by the TOV equation is smaller than $2\,\msol$.
These models can be excluded by the observed $2\,\msol$ pulsars, 
according to the entropy-dependent behavior of M-R relations~\citep{2016PhRvD..94j3001H}.

\begin{figure*}
  \centering
  \includegraphics[width=.49\linewidth]{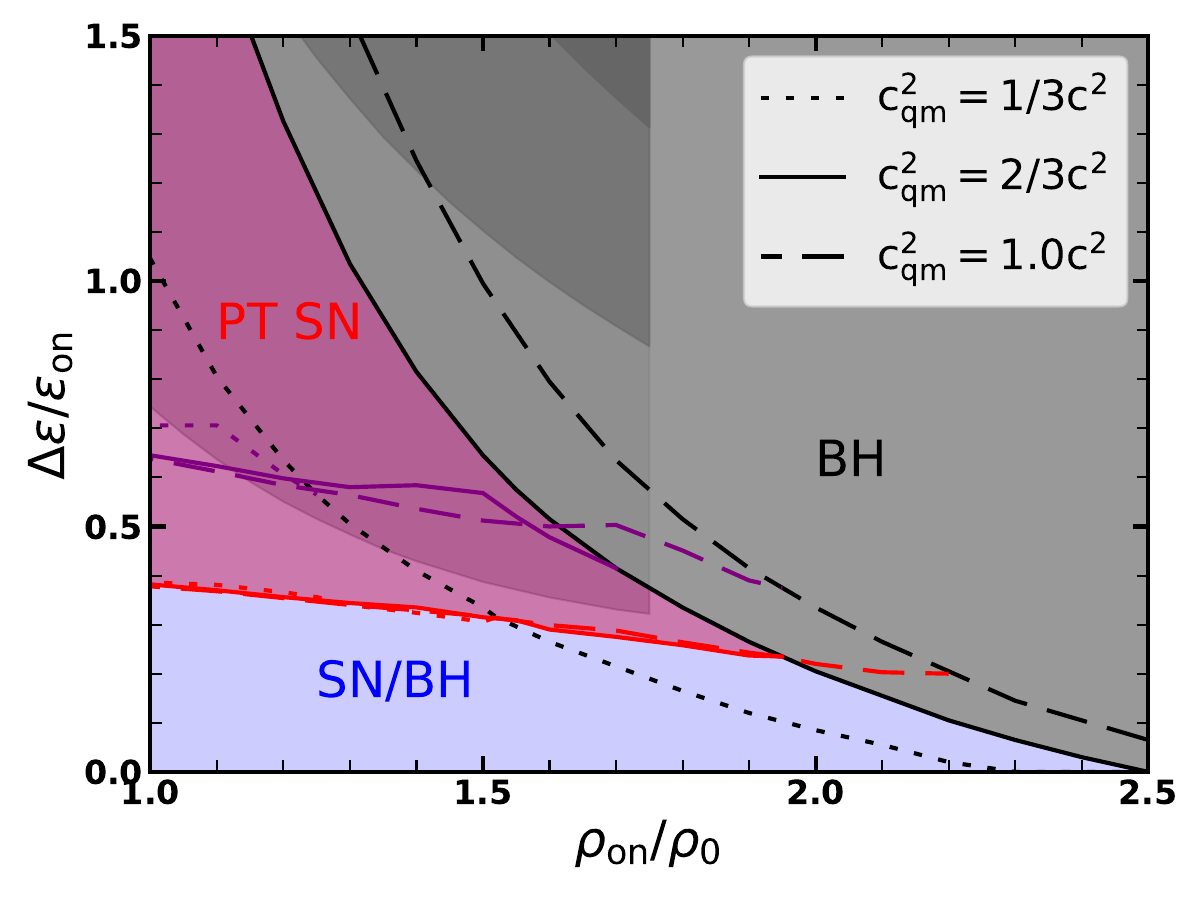}
  \includegraphics[width=.49\linewidth]{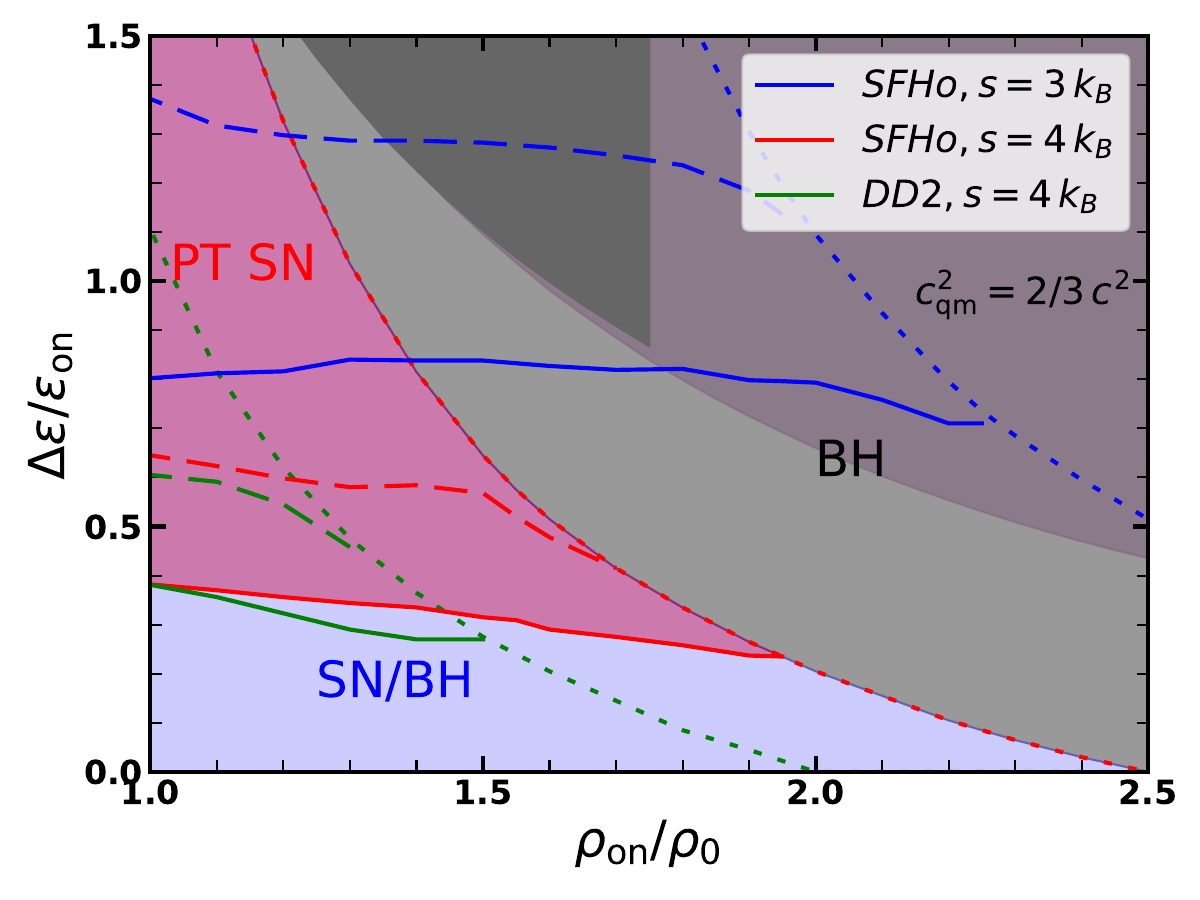}
\caption{Similar to Figure~\ref{fig:PhaseDiag1}, but showing dependence on $c^2_{\qm}$, $s$, and hadronic EoS. 
{\tt \string Left panel:} The diagrams for different quark matter speeds of sound, with $c^2_{\qm} = 1/3\,c^2$ (dotted lines), $c^2_{\qm} = 2/3\,c^2$ (solid lines) and $c^2_{\qm} = 1.0\,c^2$ (dashed lines). The SFHo EoS at $s = 4\,\kbpb$ is employed.
The critical lines of $E_{\rm exp,diag} = 0.1\,\rm B$ overlap with each other.
The shadowed area, which is eliminated by the observed 2\,$M_\sun$ pulsars, extends to a smaller $\Delta\varepsilon/\varepsilon_{\on}$ from $c^2_{\qm} = 1.0\,c^2$ to $c^2_{\qm} = 1/3\,c^2$. {\tt \string Right panel:} The diagrams for different PCS specific entropies, with $s = 3\,\kbpb$ (blue) and $s = 4\,\kbpb$ (red), and for the DD2 hadronic EoS with $s = 4\,\kbpb$ (green). $c^2_{\qm}$ is fixed at $2/3\,c^2$. The eliminated area by the observed 2\,$M_\sun$ pulsars is shaded in dark purple for $s = 3\,\kbpb$, while no area is eliminated for the case of DD2.}
\label{fig:PhaseDiag2}
\end{figure*}
The left panel of Figure~\ref{fig:PhaseDiag2} further compares the phase diagrams for quark matter with different stiffness ($c^2_{\qm}=1/3\,c^2,~2/3\,c^2$ and $1\,c^2$). 
It is found that the critical line of $E_{\rm exp,diag} = 0.1\,\rm B$ stays almost unchanged.
That of $E_{\rm exp,diag} = 1\,\rm B$ remains in the vicinity with a similar bump.
The boundary of prompt BH formation extends significantly more into the area with lower $\rho_{\on}$ and lower $\Delta\varepsilon$, once $c^2_{\qm}$ decreases.
In short, all features agree well with the findings in Figure~\ref{fig:Analy}.
In the right panel of Figure~\ref{fig:PhaseDiag2}, we test the influence of the entropy $s$.
The result shows that less $E_{\rm exp,diag}$ will be created in the collapse of a colder PCS (smaller $s$), and prompt BH formation is less likely.
This is consistent with the conclusion in \citet{2021ApJ...911...74Z}.
$E_{\rm exp,diag}$ has an upper limit at $\sim\!5\,\rm{B}$ for $s = 3\,\kbpb$.
We also test the impact of hadronic EoS with the DD2 EoS~\citep{2012ApJ...748...70H}.
In the plane of $P$ vs. $\rho$, DD2 shifts upward when compared to SFHo at $s = 4\,\kbpb$. 
The phase diagram for DD2 (see the right panel of Figure~\ref{fig:PhaseDiag2}) has a similar morphology as that of SFHo but the phase boundaries are shifted downwards.

\section{Conclusions} \label{sec:sum}
In this work, we systematically explore the impact of hadron-quark PTs on mock PCSs resembling those in the early postbounce epoch of CCSNe.
We construct a controlled series of hybrid EoSs using the CSS parameterization for quark matter and Maxwell construction for the hadron-quark mixed phase, and then perform general relativistic hydrodynamic simulations to study the outcomes of the collapse of mock isentropic PCS models.
Our idealized PCS models with a constant entropy and lepton fraction show evolutionary scenarios similar to those in detailed simulations.
We further infer the explosions of realistic CCSNe according to the diagnostic energy of the bouncing outflow.
Our study reveals the dependence of PT-induced explosions on the PT and quark matter characteristics as follows: 
\begin{itemize}
    \item The explosion is generally strengthened by a larger $\rho_{\on}$ or $\Delta \varepsilon$, while it is less sensitive to $c^2_{\qm}$.
    \item The explosion energy strongly correlates with the binding energy released as the PCS collapses, with $\log (E_{\rm exp,diag})\propto \log (\Delta E_B)$. Tentatively, the maximum $E_{\rm exp,diag}$ is $\sim\!6\,\rm{B}$ for this explosion mechanism.
    \item An increase of $c^2_{qm}$ from $1/3\,c^2$ to $1.0\,c^2$ increases the efficiency for the released binding energy transforming into the explosion energy, by a factor of $\sim$2-4.
    \item Beyond a critical $\Delta\varepsilon$ the PT leads to the BH formation instead of an explosion. At a larger $\rho_{\on}$ or smaller $c^2_{\qm}$, the critical value of $\Delta\varepsilon/\varepsilon_{\on}$ for the prompt BH formation to occur is smaller. 
    \item The morphology of the phase diagrams for the fates of PT-induced collapse of PCSs is rather generic across different $c^2_{\qm}$, PCS entropies, and hadronic EoSs. 
\end{itemize}

The above results are indicative since they are obtained under the assumption of isentropic EoSs and mock PCSs with a constant electron lepton number fraction, as well as an ``ideal-gas'' model for extra thermal effect. Such simplifications are meant to catch the main features of the PT-induced collapse of PCSs. 
Our isentropic EoSs have flat slopes of the pressures across the PT region due to the Maxwell construction, while in reality there may be a shallow slope. No additional repulsive interaction is considered in quark matter \citep{2021PhRvD.103b3001B,2024ApJ...964..143K} which results in a larger radius of the quark core and thus reduces the released binding energy during the PCS collapse. Our simulations cannot account for the overburden of stellar envelopes above the PCSs.
Overall, our $E_{\rm exp,diag}$ represents the upper limit of the final explosion energies in realistic supernova simulations.

Another caveat is that the properties of a hadron-quark PT generally depend on the temperature (as well as the electron fraction), and this further depends on the detailed microphysical models~\citep{2021PhRvD.103b3001B,2023PhRvD.108f3032B}. Our results only correspond to the PT characteristics at a finite entropy and $\beta$-equilibrium with a given electron lepton fraction. These effects are beyond the capabilities of our idealized PCS models and hybrid EOSs, highlighting the need for future developments.

Despite the caveats, the generic morphology of the phase diagrams provides useful guidance for future surveys employing realistic EoSs and detailed CCSN simulations.
Searching for hadron-quark PTs in such extreme astrophysical conditions is complementary to efforts at future experimental facilities, e.g., FAIR at GSI~\citep{2011LNP...814.....F},  NICA at Dubna~\citep{2016EPJA...52..267B} and HIAF at Huizhou~\citep{Xiaohong:2018weu},
to study high-density matter in heavy-ion collisions.

\begin{acknowledgments}
We thank the anonymous referee for the useful comments. We thank Sophia Han and Chen Zhang for the useful discussions.
X.-R.H. acknowledges support from Shanghai Jiao Tong University via the Fellowship of Outstanding PhD Graduates.
This work is partially supported by grants from the Research Grant Council of the Hong Kong Special Administrative Region, China (Project Nos. 14300320 and 14304322). S.Z. is supported by the National Natural Science Foundation of China (NSFC, Nos. 12288102, 12473031, 12393811), the Yunnan Revitalization Talent Support Program--Young Talent project, the Natural Science Foundation of Yunnan Province (No. 202201BC070003) and the Yunnan Fundamental Research Project (No. 202401BC070007). E.O. is supported by the Swedish Research Council (Project No. 2020-00452).
L.-W.C is supported by the National Natural Science Foundation of China under Grant No. 12235010, the National SKA Program of China No. 2020SKA0120300, and the Science and Technology Commission of Shanghai Municipality under Grant No. 23JC1402700.
\end{acknowledgments}

\software{GR1D \citep{2010CQGra..27k4103O,2015ApJS..219...24O},
          Numpy \citep{2020Natur.585..357H}, 
          Matplotlib \citep{2007CSE.....9...90H}
          }

\bibliography{reference}{}
\bibliographystyle{aasjournal}

\end{document}